\documentclass[reprint, amsmath,amssymb, aps, floatfix]{revtex4-2}

\usepackage{graphicx}
\usepackage{dcolumn}
\usepackage{bm}
\usepackage{hyperref}

\begin{document}


\title{X17 discovery potential from $\gamma D \to e^+ e^- pn$ with neutron tagging}

\author{Cornelis J.G. Mommers}
\email{cmommers@uni-mainz.de}
\author{Marc Vanderhaeghen}
\affiliation{Institut für Kernphysik and PRISMA${}^+$ Cluster of Excellence, \\ Johannes Gutenberg-Universität, D-55099 Mainz, Germany}

\date{\today}

\begin{abstract}
We propose a novel direct search experiment for X17 using the reaction $\gamma D \to e^+ e^- pn$. X17 is a hypothetical particle conjectured by the ATOMKI collaboration to explain anomalous signals around 17 MeV in excited ${}^8$Be, ${}^4$He and ${}^{12}$C nuclear decays via internal pair creation. It has been subject to a global experimental and theoretical research program. The proposed direct search in $\gamma D \to e^+ e^- pn$ can verify the existence of X17 through the production on a quasi-free neutron, and  determine its quantum numbers separate from ongoing and planned nuclear-decay experiments. This is especially timely in view of the theoretical tension between results from the ${}^{12}$C and ${}^8$Be measurements. Using the plane-wave impulse approximation, we quantify the expected signal and background for pseudoscalar, vector and axial-vector X17 scenarios. We optimize the kinematics for the quasi-free neutron region with the upcoming MAGIX experiment at MESA in mind and show that for all three scenarios the X17 signal is clearly visible above the QED background.

\end{abstract}

\maketitle


Despite its amazing success, the Standard Model of particle physics is incomplete. For example, it fails to account for dark matter, neutrino oscillations and the strong CP problem. This has led to ongoing research into improvements and extensions `beyond the Standard Model'. Chief among these searches is the inclusion of undiscovered particles. With heavier particles either excluded or experimentally out of reach, a vigorous effort is presently underway to search for light dark sector particles in the MeV - GeV mass range \cite{Beacham:2019nyx,Agrawal:2021dbo}. Recent results of the ATOMKI collaboration have garnered significant theoretical and experimental interest; in a series of experiments \cite{Krasznahorkay:2015ijz, Krasznahorkay:2021joi,Krasznahorkay:2022pxs} the collaboration claims to have found evidence of a new, light boson dubbed X17.

The ATOMKI collaboration looked at internal pair creation in decays of excited ${}^8$Be, ${}^4$He and, recently, ${}^{12}$C nuclei. In all three cases, an anomalous bump was found in the distribution of the emitted electron-positron pair's relative angle, with a statistical significance consistently exceeding $6 \sigma$ (see Ref.~\cite{Alves:2023ree} for a review). In the Standard Model, nuclear transitions where an $e^+ e^-$ pair is emitted are mediated by electromagnetic interactions and are well understood. They are sensitive to new physics appearing at the MeV scale, and thus the ATOMKI collaboration attributes their anomaly to the as-of-yet unseen X17, with a reported averaged mass of $17.02(10)$ MeV \cite{Krasznahorkay:2022pxs,Krasznahorkay:2021joi, Krasznahorkay:2015ijz}. Assuming definite parity, the beryllium results indicate X17 can be a pseudoscalar, vector or axial-vector particle \cite{Krasznahorkay:2015ijz}, while the carbon results point to a scalar, vector or axial-vector particle \cite{Barducci:2022lqd}. Theoretically, models for X17 have been developed for the pseudoscalar, vector and axial-vector cases \cite{Feng:2016ysn,Ellwanger:2016wfe,Kozaczuk:2016nma,Alves:2017avw,Viviani:2021stx} that can explain the ATOMKI anomalies while conforming to existing exclusion bounds. In particular, according to the vector model put forward by Feng \textit{et al}. \cite{Feng:2016jff,Feng:2016ysn} X17 must be protophobic (couple weakly to protons) to meet existing bounds from the NA48/2 experiment \cite{NA482:2015wmo}.  Experimentally, a global effort is underway to scrutinize the results of the ATOMKI anomaly, with new experiments such as CCPAC \cite{Azuelos:2022nbu}, MEG II at the PSI \cite{MEGII:2018kmf} among others \cite{Alves:2023ree}.

Many of these ongoing experiments focus on nuclear decays. After all, this is where X17 was first observed. However, if X17 is a bona fide new particle, then it must also play a role in other processes. Assuming the size of the X17 couplings needed to explain the ATOMKI anomaly, it was estimated that contributions from X17 to the reaction $\gamma n \to e^+ e^- n$ would be clearly visible over the QED background. The latter is suppressed for a neutron target \cite{Backens:2021qkv}, enabling a direct search for X17 at electron accelerators. The upcoming MAGIX experiment at MESA \cite{Doria:2019sux} is ideal for this, due to MESA's low-energy yet high-intensity electron beam (105 MeV in its energy-recovering mode) and MAGIX’s high-resolution spectrometers, capable of resolving the invariant mass of the outgoing $e^+ e^-$  pair to at least 0.1 MeV.

Of course, in the lab one does not have access to a free, high-density neutron target, so processes like $\gamma n \to e^+ e^- n$ are not directly measurable. Instead, in this work we propose a novel direct search experiment using neutron tagging \cite{Levchuk:1994ij} with dilepton photoproduction on a deuteron, $\gamma D \to e^+ e^- pn$. By tagging the neutron we can treat the bound neutron as quasi-free and the bound proton as a spectator. In this way, scattering events take place primarily on the `nearly on-shell' quasi-free neutron. We investigate the X17 signal relative to the QED background for the pseudoscalar, vector and axial-vector X17 scenarios in such an experiment, for a kinematic regime accessible by the MAGIX experiment at MESA.

\begin{figure}[t]
    \centering
    \includegraphics[width=8.6cm]{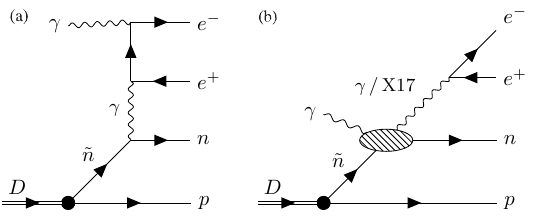}
    \caption{Tree-level diagrams for the reaction $\gamma D \to e^+ e^- pn$ within the plane-wave impulse approximation. Crossed diagrams as well as diagrams in which the bound neutron is a spectator are omitted. (a) The Bethe-Heitler process. (b) The Compton and X17 production processes on a quasi-free neutron.}
    \label{fig:diagrams}
\end{figure}

Let us begin with specifying the kinematics. All quantities are given in the lab frame in which the deuteron is at rest. We consider the reaction
\begin{align}
    &\gamma(E_\gamma, \mathbf{q}, \lambda) \, D(m_D, \mathbf{0}, M) \to e^+(E_+, \mathbf{p}_+, s_+)\nonumber \\
    &\qquad \quad e^-(E_-, \mathbf{p}_-, s_-) \, p(E_p, \mathbf{p}_p, s_p) \, n(E_n, \mathbf{p}_n, s_n),
\end{align}
where $\lambda$ is the polarization of the incoming photon, and $s_\pm$, $s_p$ and $s_n$ are $e^\pm$, $p$ and $n$ helicities, respectively, and $M$ is the deuteron spin projection on the $z$-axis. The $z$-axis is chosen along the direction of the photon momentum. The masses of the nucleons and deuteron are denoted by $m_N$ and $m_D$, respectively. The momenta of the virtual photon or X17 is given by $q' = p_+ + p_-$ and we denote the invariant mass of the dilepton system by $q'^2 = m_{ee}^2$. Our kinematic variables are $E_\gamma$, $|\mathbf{p}_\pm|$, the polar angles $\theta_\pm$ and $\theta_n$, and the azimuthal angles $\phi_\pm$ and $\phi_n$. All polar angles are defined with respect to the $z$-axis. Lastly, the cross section is given by
 \begin{align}
    \frac{ \mathrm{d} \sigma }{\mathrm{d}\Pi} &=  \frac{1}{ 64 ( 2\pi )^{8} m_D E_\gamma }  \frac{ | \mathbf{p}_+ |^2 | \mathbf{p}_- |^2 }{E_+ E_-} \nonumber \\
    &\quad \times  \frac{ | \mathbf{p}_n |^2 }{| \mathbf{p}_n | \left( m_D + E_\gamma - q'^0 \right) - E_n | \mathbf{q} - \mathbf{q}' | \cos\theta_{n\gamma\gamma} } \nonumber \\
    &\quad \times  \langle | \mathcal{M} |^2 \rangle,
 \end{align}
where $\mathrm{d}\Pi$ is shorthand for $ \mathrm{d} |\mathbf{p}_+| \mathrm{d} | \mathbf{p}_- | \mathrm{d} \Omega_n \mathrm{d} \Omega_-  \mathrm{d} \Omega_+ $. Here, $\langle \lvert \mathcal{M} \rvert^2 \rangle$ is the spin-averaged Feynman matrix element and $\theta_{n\gamma\gamma}$ is the angle between $\mathbf{p}_n$ and $\mathbf{q} - \mathbf{q}'$. 

To calculate $\mathcal{M}$ we use the plane-wave impulse approximation (PWIA). This allows us to separate the process $\gamma D \to e^+ e^- pn$ into two parts: a part where the proton is a spectator and another where the neutron is a spectator. By doing so, we disregard meson exchange currents and final state interactions. However, in our kinematic regime of interest the meson exchange currents are estimated to give corrections of approximately 5\%, meaning they can be safely neglected. Likewise, for a first approximation, the final state interactions can be omitted (see Fig.~5 and its discussion in Ref.~\cite{Levchuk:1994ij}. We have checked our model against the solid line in Fig.~5 and found both results to agree reasonably well). 

In the PWIA, the quasi-free neutron amplitude is given by
\begin{align}
    &\mathcal{M}_{n}^\text{quasi-free} \left( \gamma \, D \to e^+ \, e^- \, p \, n \right) = (2 m_D)^{1/2} \left(\frac{E_p}{E_{\tilde{n}}}\right)^{1/2} \nonumber \\
    &\quad \times  \sum_{s_{\tilde{n}}} \tilde{\Psi}^{M}_{s_p s_{\tilde{n}}} \left( \mathbf{p}_p \right) \mathcal{M}\left( \gamma \, \tilde{n} \to e^+ \, e^- \, n \right) ,
 \end{align}
where the quasi-free neutron, $\tilde{n}$, has momentum and spin projection $-\mathbf{p}_p$ and $s_{\tilde{n}}$, respectively. The relative deuteron wave function in momentum space is
\begin{align}
    &\tilde{\Psi}^M_{s_p s_{\tilde{n}}}(\mathbf{p}_p) = (2\pi)^{3/2} \bigg\{ \frac{1}{\sqrt{4\pi}} \tilde{\psi}_0(|\mathbf{p}_p|) \langle \tfrac{1}{2} s_p; \tfrac{1}{2} s_{\tilde{n}} \vert 1 M \rangle  \nonumber \\
    &\quad - \tilde{\psi}_2(|\mathbf{p}_p|)\sum_{M_s} Y_{2(M - M_s)}(\hat{\mathbf{p}}_p) \langle 1 M_s; 2 M - M_s \vert 1 M \rangle \nonumber \\
    &\qquad \qquad \qquad \quad \times \langle \tfrac{1}{2} s_p; \tfrac{1}{2} s_{\tilde{n}} \vert 1 M_s \rangle  \bigg\},
\end{align}
where $\langle j_1 m_1; \, j_2 m_2 \vert j m \rangle$ are the Clebsch-Gordan coefficients and $Y_{LM}$ are the spherical harmonics. We use the CD-Bonn parametrization \cite{Machleidt:2000ge} for the s- and d-wave functions, $\tilde{\psi}_0$ and $\tilde{\psi}_2$ respectively.

The relevant diagrams are given in Fig.~\ref{fig:diagrams}. The QED background processes include the Bethe-Heitler process (a) and Compton scattering (b). For the energy range of the MAGIX experiment at MESA, with $E_\gamma$ around 100 MeV, we can describe the Compton amplitude as the sum of Born, $\pi^0$ $t$-channel exchange, and electric ($\alpha_E$) and magnetic ($\beta_M$) nucleon polarizability contributions. The latter are parameterized by a low-energy expansion as detailed in Ref.~\cite{Lensky:2017bwi}. The main contribution to the X17 signal process comes from the Born amplitude in Fig.~\ref{fig:diagrams}(b), in which X17 is produced on a nucleon. In the Bethe-Heitler process, a possible X17 contribution is far off-resonance and is therefore negligible.

To estimate the coupling of X17 to the nucleon we employ models by Alves and Weiner \cite{Alves:2017avw} for the pseudoscalar case, by Feng \textit{et al}. \cite{Feng:2016ysn} for the vector case and by Kozaczuk \textit{et al}. \cite{Kozaczuk:2016nma} for the axial-vector case. The relevant Lagrangians are
\begin{align}
    \mathcal{L}_\text{P} &= i \bar{N} \gamma_5 \left( g^{(0)}_{XNN} + g^{(1)}_{XNN} \tau_3 \right) N X,\\
    \mathcal{L}_\text{V} &= -e X_\mu \sum_{N = p,n} \varepsilon_N \bar{N} \gamma^\mu N,\\
    \mathcal{L}_\text{A} &= - X_\mu \sum_{N = p,n} a_N \bar{N} \gamma^\mu \gamma_5 N,
\end{align}
where $\tau_3$ is the isospin Pauli matrix, $g^{(0)}_{XNN}$ and $g^{(1)}_{XNN}$ the isoscalar and isovector pseudoscalar couplings, respectively, $\varepsilon_{p,n}$ the vector couplings, $e>0$ the proton charge and $a_{p,n}$ the axial-vector couplings.

\begin{figure}[t]
    \centering
    \includegraphics[width=8.6cm]{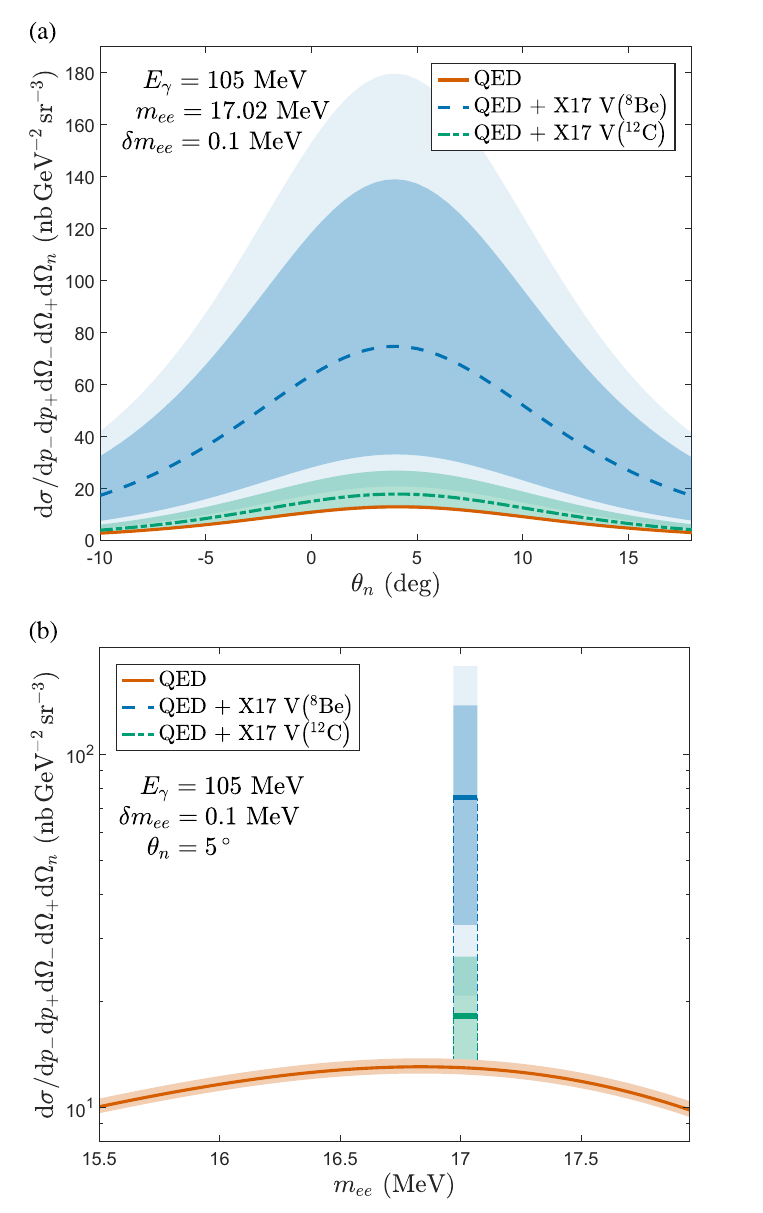}
    \caption{The differential cross section for the reaction $\gamma D \to e^+ e^- p n$ around the neutron quasi-free peak for in-plane kinematics (see text) as it would appear in a single bin with a width of $0.1$ MeV around $m_{ee} = 17.02$ MeV (a) or in a histogram during a bump hunt (b). The red line shows the sum of the QED processes, while the dashed blue line and dash-dotted green line show the sum of the QED and vector (V) X17 signal. The blue and green lines indicate couplings derived from the ${}^8$Be \cite{Krasznahorkay:2015ijz} or ${}^{12}$C experiments \cite{Krasznahorkay:2022pxs}, respectively. Dark and light bands indicate a $2\sigma$ and $3\sigma$ uncertainty range in the couplings. The red band indicates uncertainties in the QED calculation due to the neutron polarizabilities, taken from the PDG~\cite{Workman:2022ynf}.}
    \label{fig:cs_V}
\end{figure}

We can constrain X17's couplings by using the branching fractions of the ${}^8$Be and ${}^{12}$C decays reported by the ATOMKI collaboration \cite{Krasznahorkay:2015ijz,Krasznahorkay:2022pxs},
\begin{align}
    \frac{\Gamma_X}{\Gamma_\gamma} \bigg\vert_{{}^8\text{Be}(18.15)} &= 6(1) \times 10^{-6},\\
    \frac{\Gamma_X}{\Gamma_\gamma} \bigg\vert_{{}^{12}\text{C}(17.23)} &= 3.6(3) \times 10^{-6},
\end{align}
which can be translated to limits on the X17 couplings \cite{Barducci:2022lqd}. We assume that the electronic branching fraction, $\mathcal{B}(X \to e^+ e^-)$, which always appears as an overall factor, is equal to unity. Note that the ${}^8$Be(18.15) state, which is predominately isoscalar, is isospin mixed with the ${}^8$Be(17.64) state, which is predominately isovector. In our multipole analysis we parameterize this isospin mixing with an isospin-mixing angle, $\theta_{1^+}$, and an isospin-breaking parameter, $\kappa$ \cite{Feng:2016ysn}. Following Ref.~\cite{Alves:2017avw}, we take $\theta_{1^+} = 0.35(8) {}^\circ$, whence $\kappa = 0.681$ \cite{Feng:2016ysn}. 

For a pseudoscalar X17 scenario, results from the SINDRUM collaboration \cite{SINDRUM:1986klz} put a strong bound on the isovector coupling, $|g^{(1)}_{XNN}| \leq 0.6 \times 10^{-3}$. By following the procedure described in Ref.~\cite{Alves:2017avw} we derive a limit on the isoscalar coupling. 

For a vector X17 scenario the constraint provided by the NA48/2 experiment \cite{NA482:2015wmo} leads to the protophobia condition, $|\varepsilon_p| \leq 1.2 \times 10^{-3}$. We derive the remaining neutron coupling from the ${}^8$Be data as outlined in Ref.~\cite{Feng:2016ysn}, and from the ${}^{12}$C data using \cite{Barducci:2022lqd}
\begin{equation}
    \frac{\Gamma_X}{\Gamma_\gamma} \bigg\vert_{{}^{12}\text{C}(17.23)} = \frac{k}{\Delta E} \left( 1 + \frac{m_X^2}{2 \Delta E^2} \right) | \varepsilon_p - \varepsilon_n |^2,
\end{equation}
where $k = \sqrt{\Delta E^2  - m_X^2}$.  

To derive couplings for the scenario of an axial-vector X17 we need its nuclear matrix elements. For the carbon transition these matrix elements are unknown, and computing them is outside the scope of this work. For the beryllium transition we take the matrix elements as calculated in Ref.~\cite{Kozaczuk:2016nma},
\begin{align*}
    \langle {}^8\text{Be}(\text{g.s.}) \vert \vert \hat{\sigma}^{(p)} \vert \vert {}^8\text{Be}(18.15) \rangle &= (-0.38 \pm 2.19) \times 10^{-2}, \\
    \langle {}^8\text{Be}(\text{g.s.}) \vert \vert \hat{\sigma}^{(n)} \vert \vert {}^8\text{Be}(18.15) \rangle &= (-10 \pm 2.6) \times 10^{-2}.
\end{align*}

Our coupling values are presented in Table \ref{tab:coupling}. As noted in Ref.~\cite{Barducci:2022lqd}, there is some tension between the carbon and beryllium results for a vector-like X17. The neutron couplings only overlap when the uncertainty of the ${}^8$Be results is increased to around 3$\sigma$. Indeed, this discrepancy highlights the need for independent verification of X17, as proposed here. 

\begin{table}[!b]
    \caption{Values for X17's couplings to the nucleon for the scenarios where X17 is a pseudoscalar (P), vector (V) or axial-vector (A) particle. Coupling ranges include a $1\sigma$ variation around the central value.}
    \begin{tabular}{l| r c@{$\null=\null$} l | l}
        \hline
        \hline
        Parity & \multicolumn{3}{c|}{Value coupling} & Ref.\\      
        \hline
        P$\left({}^8\text{Be}\right)$ & $|g^{(0)}_{XNN}|$ && $(2.0 - 5.4) \times 10^{-3}$ & \cite{Alves:2017avw}\\
        & $|g^{(1)}_{XNN}|$ && $(0.0 - 0.6) \times 10^{-3}$ & \cite{SINDRUM:1986klz}\\
        V & $|\varepsilon_p|$ && $(0.0 - 1.2) \times 10^{-3}$ & \cite{NA482:2015wmo}\\
        V$\left({}^8\text{Be}\right)$ & $|\varepsilon_n|$ && $(1.1 - 1.7) \times 10^{-2}$ & \cite{Feng:2016ysn,Feng:2016jff}\\
        V$\left({}^{12}\text{C}\right)$ & $|\varepsilon_n|$ && $(2.6 - 5.3) \times 10^{-3}$ & \cite{Barducci:2022lqd}\\
        A$\left({}^8\text{Be}\right)$ & $|a_{p,n}|$ && $(2.7 - 11.6) \times 10^{-5}$ & \cite{Kozaczuk:2016nma}\\ 
        \hline
        \hline
    \end{tabular}
    \label{tab:coupling}
\end{table}

\begin{figure}[t]
    \centering
    \includegraphics[width=8.6cm]{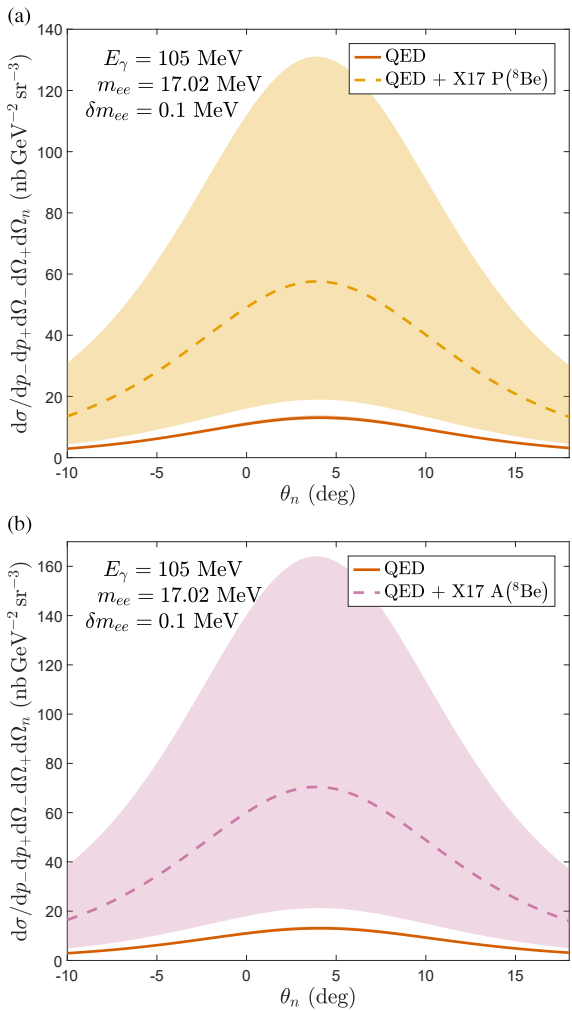}
    \caption{The differential cross section for the reaction $\gamma D \to e^+ e^- p n$ around the neutron quasi-free peak using in-plane kinematics (see text). The red line represents the sum of all QED background processes, while the dashed yellow (a) and dashed magenta (b) lines show the sum of the QED background and pseudoscalar (P) or axial-vector (A) X17 signal, respectively. The signal is averaged over a 0.1 MeV bin around $m_{ee} = 17.02$ MeV. The bands indicate a $1\sigma$ uncertainty range in the X17 coupling to the nucleon.}
    \label{fig:cs_PA}
\end{figure}

To maximize the signal-to-background ratio for the quasi-free neutron contribution to the $\gamma D \to e^+ e^- pn$ process, we must optimize the kinematics. In doing so, to ensure the validity of the quasi-free neutron process in the PWIA, we have to remain within the neutron quasi-free peak (NQFP) region, defined by \cite{Levchuk:1994ij}
\begin{equation}
    | \mathbf{p}_p | \lesssim \sqrt{m_N \Delta} \approx 45.7 \, \text{MeV/c},
\end{equation}
where $\Delta \approx 2.2$ MeV is the deuteron binding energy. Our kinematics are optimized for the MAGIX experiment at MESA with a beam energy of $E_\gamma = 105$ MeV and we consider in-plane kinematics where $\phi_\pm = 0.0 \, {}^\circ$ and $\phi_n = 0.0 \, {}^\circ$. We found the optimal kinematics, consistent with the above, to be an asymmetric backward configuration for the lepton pair,
\begin{align*}
    &|\mathbf{p}_+| = 65.7 \text{ MeV/c}, \quad \theta_+ = -165.0 \, {}^\circ, \\
    &|\mathbf{p}_-| = 20.1 \text{ MeV/c}, \quad \theta_- = 168.0 \, {}^\circ, \\
    &\theta_n = 5.0 \, {}^\circ,
\end{align*}
where negative angles indicate that the positron is emitted in the opposite half plane in comparison with the electron and neutron. The NQFP corresponds to an angular range $\theta_n \in \left[ -10, 18 \right] \, {}^\circ$.

Figure \ref{fig:cs_V} shows the quasi-free neutron cross section for a vector X17 as a function of $\theta_n$ or $m_{ee}$. The blue and green signals are derived from beryllium and carbon, respectively. The red QED background includes uncertainties from neutron polarizabilities as taken from the PDG \cite{Workman:2022ynf}. Dark and light bands represent 2$\sigma$ and 3$\sigma$ variations in X17’s couplings. Figures \ref{fig:cs_PA}(a) and \ref{fig:cs_PA}(b) depict the pseudoscalar and axial-vector scenarios in yellow and magenta with a $1\sigma$ variation in the coupling. The signal cross section is averaged over a bin of $\delta m_{ee} = 0.1$ MeV, corresponding to the expected resolution of MAGIX.

In Fig.~\ref{fig:cs_V}(a) we see that the signal is visible above the QED background for both the beryllium- or carbon-derived couplings. In Fig.~\ref{fig:cs_V}(b) we see that in a bump-hunt-style search the presence of X17 would cause a spike in a single bin, which would be particularly noticeable if we use the neutron-coupling values derived from the ${}^8$Be experiment. A signal is also expected for the other two scenarios in Fig.~\ref{fig:cs_PA}, where the X17 signal juts out above the QED background.

In summary, we have studied the reaction $\gamma D \to e^+ e^- pn$ with quasi-free neutron kinematics in the context of a novel, direct search for the X17 particle conjectured by the ATOMKI collaboration. Using the plane-wave impulse approximation we have shown that the X17 signal is visible above the QED background for a pseudoscalar, vector and axial-vector X17 scenario. Furthermore, we have shown that such a process can be experimentally accessed at the upcoming MAGIX experiment at MESA. Due to the uncertainty surrounding X17’s nature and the potential tension between the carbon and beryllium results, an experiment at an electron scattering facility like MESA, with its unparalleled $e^+ e^-$ invariant-mass resolution, will  provide an important and timely check of ongoing nuclear-decay experiments. 

The authors thank S. Schlimme for helpful communications. This work was supported by the Deutsche Forschungsgemeinschaft (DFG, German Research Foundation), in part through the Research Unit [Photon-photon interactions in the Standard Model and beyond, Projektnummer 458854507 - FOR 5327], and in part through the Cluster of Excellence [Precision Physics, Fundamental Interactions, and Structure of Matter] (PRISMA$^+$ EXC 2118/1) within the German Excellence Strategy (Project ID 39083149).

\bibliography{bib.bib}

\end{document}